\documentclass[twocolumn,preprintnumbers,secnumarabic,amsmath,amssymb,groupedaddress,nofootinbib]{revtex4-2}

\pdfoutput=1
\usepackage{amssymb,amsmath,latexsym,mathrsfs}
\usepackage{url}
\usepackage{enumitem}
\usepackage{bm}
\usepackage{graphicx}
\usepackage{booktabs}
\usepackage[usenames,dvipsnames]{color}
\usepackage[breaklinks,colorlinks,urlcolor=magenta,citecolor=magenta,linkcolor=magenta]{hyperref}
\usepackage{multirow}
\usepackage{float}
\usepackage{cases}
\usepackage{blindtext}
\usepackage{pifont}
\usepackage{hhline}
\usepackage{siunitx}
\usepackage{braket}
\usepackage[table,xcdraw]{xcolor}
\definecolor{linkcolor}{rgb}{0.0, 0.47, 0.75}
\definecolor{citecolor}{rgb}{1.0, 0.5, 0.0}
\definecolor{linkcolor}{rgb}{0.390625,0.5607843137,0.99609375}
\hypersetup{
  linkcolor  = linkcolor,
  citecolor  = linkcolor,
  urlcolor   = linkcolor,
  colorlinks = true
}
\hypersetup{
  linkcolor  = linkcolor,
  citecolor  = linkcolor,
  urlcolor   = linkcolor,
  colorlinks = true
}
\usepackage{verbatim}
\usepackage{hyperref}
\usepackage[normalem]{ulem}
\usepackage{accents}
\usepackage{tikz}
\newcommand\ignore[1]{}			% Use pdf, png, jpg, or eps§ with pdflatex; use eps in DVI mode
\usepackage{tocloft}
\usepackage{slashed}
\usepackage{float}
\usepackage{fullpage}
\usepackage{url}

\usepackage{tabularx}

\usepackage{upgreek}
\usepackage{outlines}
\def\0{{(0)}}
\def\1{{(1)}}
\usepackage{bm}

\def\ccccend{\end{array}\right)}

\usetikzlibrary{matrix,arrows,decorations.pathmorphing}
\usetikzlibrary{patterns}
\usetikzlibrary{shapes.misc}
\usetikzlibrary{trees}
\usetikzlibrary{decorations.pathmorphing}
\usetikzlibrary{shapes.geometric}
\usetikzlibrary{positioning}
\usetikzlibrary{decorations.markings}
\usetikzlibrary{positioning,arrows}

\usetikzlibrary{decorations.pathreplacing}
\usetikzlibrary{decorations.pathmorphing}
\usetikzlibrary{shapes}
\usetikzlibrary{arrows.meta}
\usetikzlibrary{plotmarks}
\usetikzlibrary{decorations.markings}
\tikzset{
particle/.style={thin,draw=black, postaction={decorate},
decoration={markings,mark=at position .5 with {\arrow[black, line width=0.5mm]{stealth}}}},
gluon/.style={decorate, draw=black, decoration={coil,amplitude=4pt, segment length=5pt}},
photon/.style={decorate, decoration={snake}},
singularity/.style={decorate, draw=black, decoration=zigzag}
}

\usepackage[bottom]{footmisc}

\usepackage{amsthm}
\theoremstyle{theorem}

\begin{document}

\title{Possible Implications of QCD Axion Dark Matter Constraints from Helioscopes \\
and Haloscopes for the String Theory Landscape } 
%\title{The QCD axion as a measure of topology of extra dimensions}
\author{Naomi Gendler$^{a}$}
\author{David J. E. Marsh$^{b}$}
\affiliation{$^{a}$Jefferson Physical Laboratory, Harvard University, Cambridge, MA 02138 USA}
\affiliation{$^{b}$Theoretical Particle Physics and Cosmology, King's College London, Strand, London, WC2R 2LS, United Kingdom}

\begin{abstract}

Laboratory experiments have the capacity to detect the QCD axion in the next decade, and precisely measure its mass, if it composes the majority of the dark matter. In type IIB string theory on Calabi-Yau threefolds in the geometric regime, the QCD axion mass, $m_a$, is strongly correlated with the topological Hodge number $h^{1,1}$. We compute $m_a$ in a scan of $185{,}965$ compactifications of type IIB string theory on toric hypersurface Calabi-Yau threefolds. We compute the range of $h^{1,1}$ probed by different experiments under the condition that the QCD axion can provide the observed dark matter density with minimal fine-tuning. Taking the experiments DMRadio, ADMX, MADMAX, and BREAD as indicative on different mass ranges, the $h^{1,1}$ distributions peak near $h^{1,1}=24.9, \ 65.4, \ 196.8,$ and $310.9,$ respectively. We furthermore conclude that, without severe fine tuning, detection of the QCD axion as dark matter \emph{at any mass} disfavours 80\% of models with $h^{1,1} =491$, which is thought to have the most known Calabi-Yau threefolds. Measurement of the solar axion mass with IAXO is the dominant probe of all models with $h^{1,1}\gtrsim 250$. This Letter demonstrates the immense importance of axion detection in experimentally constraining the string landscape. %\tkDM{IAXO comment for 491? May not fit.} \nvg{To me doesn't seem to fit in abstract, maybe in intro?} I think it's best left for discussion section not intro.

% Resonator-based experiments offer the most promising avenue for detecting the QCD axion if it comprises all of the dark matter in our universe. Different resonators probe different mass ranges for the QCD axion, from XX to XX eV. \nvg{link between fs and topology} We compute the mass of the QCD axion in a scan of XXX compactifications of type IIB string theory on toric hypersurface Calabi-Yau threefolds. We present the favored range of $h^{1,1}$ in the scenario that each resonator experiment were to detect the QCD axion as dark matter. The distributions peak at $h^{1,1} = XX$ in the case of an DM Radio \cite{} detection, $h^{1,1} = XX$ in the case of an ADMX \cite{} detection, and $h^{1,1} = XX$ in the case of a MADMAX \cite{} detection. We furthermore conclude that any detection of the QCD axion as dark matter rules out the vast majority of models in this landscape with $h^{1,1} =491$ in the geometric regime.
\end{abstract}

\maketitle

\textit{Introduction}---The quantum chromodynamics (QCD) axion is a hypothetical particle, $\theta$, first proposed in the 1970s~\cite{Peccei:1977hh,Weinberg:1977ma,Wilczek:1977pj} to solve the problem of CP invariance of the strong nuclear force~\cite{Abel:2020pzs}. In the early 1980s, experimentally viable versions of this model were constructed in quantum field theory~\cite{Kim:1979if,Shifman:1979if,Dine:1981rt,Zhitnitsky:1980tq}, and it was soon realized~\cite{Dine:1982ah,Preskill:1982cy,Abbott:1982af} that these ``invisible" axion models can provide an explanation for the observed~\cite{Planck:2018vyg} cosmic dark matter (DM). The QCD axion mass, $m_a$, is determined by its ``decay constant", $f_a$: $m_a = 5.70(1) \,\mu\text{eV} (10^{12}\text{ GeV}/f_a)$~\cite{Weinberg:1977ma,Wilczek:1977pj,GrillidiCortona:2015jxo}.

The QCD axion is typically searched for by its coupling to electromagnetism, $g_{a\gamma}=C \alpha/(2\pi f_a)$, where $\alpha$ is the fine structure constant, and $\mathcal{C}\sim\mathcal{O}(1)$ is model dependent~\cite{DiLuzio:2016sbl,DiLuzio:2020wdo,Plakkot:2021xyx}. Because of the immense experimental challenge to detect the axion~\cite{Wuensch:1989sa,PhysRevD.42.1297,PhysRevLett.59.839,Hagmann:1996qd}, QCD axion DM is in the relatively unique position that the original theories from the 1980's remain largely unconstrained. %, and thus experimntally and theoretically viable.} %QCD axion DM is in the relatively unique position that the axion theories have not been under pressure, and therefore barely changed in 40 years due to the immense experimental challenge to detect the axion~\cite{Wuensch:1989sa,PhysRevD.42.1297,PhysRevLett.59.839,Hagmann:1996qd}.
%QCD axion DM is in the relatively unique position that the theory goalposts have barely changed in 40 years due to the immense experimental challenge to detect the axion~\cite{Wuensch:1989sa,PhysRevD.42.1297,PhysRevLett.59.839,Hagmann:1996qd}.

The experimental landscape, however, is undergoing a sea change. In 2018 the Axion Dark Matter eXperiment (ADMX) demonstrated for the first time experimental sensitivity to the 1980s target axion DM models~\cite{Dine:1981rt,Zhitnitsky:1980tq} of a given $\mathcal{C}$, at axion mass $m_a\approx 2.7\,\mu\text{eV}$~\cite{ADMX:2018gho}. ADMX uses a resonant microwave cavity to search for axion DM~\cite{Sikivie:1983ip}, giving the ability to precisely determine the axion mass in the event of a detection. Cavity resonators, due to limitations imposed by volume at low and high frequency, are only a viable technology to detect axions over around a decade in mass~\cite{Semertzidis:2021rxs,Stern:2016bbw}.  In parallel, theoretical and experimental groups around the world have developed a wide variety of complementary technologies that have the capability to detect QCD axion DM across a much wider range, $10^{-11}\text{ eV}\lesssim m_a\lesssim 10^{-2} \text{ eV}$, within the coming decade or so~\cite{Chadha-Day:2021szb,Semertzidis:2021rxs,Adams:2022pbo}. The question, then, is: what is the axion mass? 

There are two possibilities for QCD axion DM production in the early Universe: the ``pre-inflation" and the ``post-inflation" scenarios (see reviews~\cite{Chadha-Day:2021szb,Marsh:2015xka,OHare:2024nmr}). In the pre-inflation scenario, cosmic inflation (or any alternative initial conditions scenario) has the effect of hiding from view the high energy physics responsible for the existence of the axion, leaving a uniform field everywhere at early times. There is a free (stochastic) parameter associated to this mechanism, the ``initial misalignment angle", $\theta_i$, and therefore the axion mass is able to span a wide range and still provide the correct DM relic density, $\Omega_a h^2\approx 0.12$~\cite{Planck:2018vyg}, restricted only by the fine tuning allowed on $\theta_i$. In the alternative post-inflation scenario the axion is assumed to be given by the phase of a complex field, $\Phi=\rho e^{i\theta}$, which undergoes spontaneous symmetry breaking and forms cosmic strings~\cite{Kibble:1976sj} that subsequently decay producing axions~\cite{Hagmann:1990mj,Battye:1994au}. Current numerical simulations of this scenario predict $95\,\mu\text{eV}\lesssim m_a\lesssim 450\,\mu\text{eV}$ if the axion provides the DM relic density~\cite{Gorghetto:2020qws,Buschmann:2021sdq,Saikawa:2024bta,Kim:2024wku}.
% In this complex, non-linear process, the requirement of matching $\Omega_a h^2\approx 0.12$ fixes, in principle, the axion mass.

What does string theory have to say? String theory provides a natural mechanism for the existence of the QCD axion, and generically predicts a range of other axionlike particles~\cite{Witten:1984dg,Svrcek:2006yi,Conlon:2006tq,Arvanitaki:2009fg,Cicoli:2012sz,Demirtas:2018akl,Demirtas:2021gsq}. In this context, the decay constants of the axions are typically fixed by the volumes of some internal regions of the six-dimensional compact manifold~\cite{Reece:2024wrn}, and so, in principle, axion detection can probe extra dimensions and the string theory landscape. To make concrete progress in this direction we restrict our search space to type IIB string theory compactified on Calabi-Yau (CY) threefolds~\cite{Candelas:1985en,Greene:1996cy} in the geometric regime realized as hypersurfaces in toric varieties~\cite{batyrev1993dual}. The QCD axion arises as the integral of the ten-dimensional $C_4$ potential over a four-cycle in the geometry that hosts QCD, modeled by a stack of D7 branes. The geometry is chosen such that the gauge coupling of QCD at low energy matches observations. %We return to this discussion at the end of this \emph{Letter}. \nvg{letter?} \tkDM{I've seen people write like this in PRL before, which I assume we're going for?}

\begin{figure}
    \centering
    \includegraphics[width=1.2\linewidth]{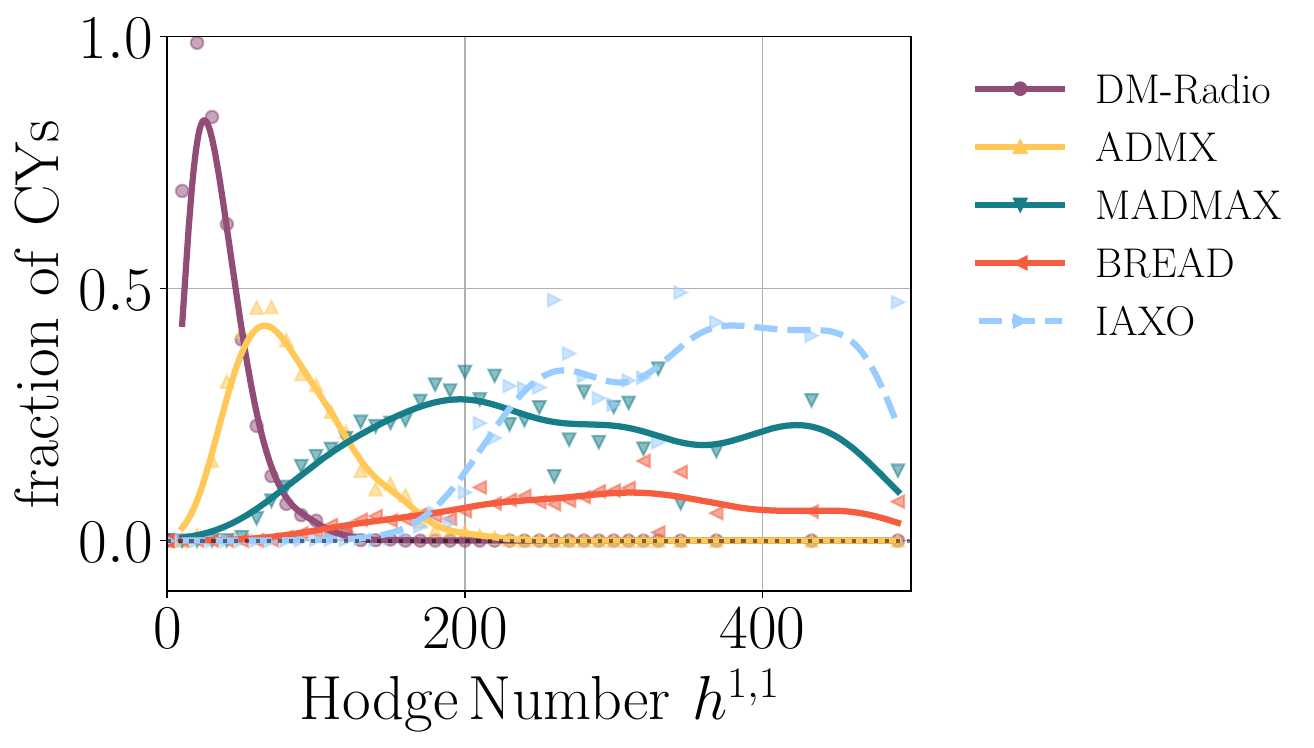}
    \caption{The fraction of Calabi-Yau threefolds that fall into the range of sensitivity for each QCD axion experiment, as a function of $h^{1,1}$.}%For $m_a$ predicted at the largest $h^{1,1}=491$ it is not possible to provide the relic density with our fine tuning penalty, so only the IAXO helioscope has sensitivity here.}
    \label{fig:h11predictions}
\end{figure}

It has been observed by us and in previous studies~\cite{Demirtas:2021gsq,Gendler:2023kjt,Demirtas:2018akl, Mehta:2021pwf} that the axion decay constant and mass in such setups is strongly correlated with the topological Hodge number $h^{1,1}$ of the CY. On the other hand, we note that such a setup alone does not uniquely predict the value of $C$ in the axion-photon coupling, since this depends on the charged matter content below the compactification scale. String theory predicts the QCD axion mass in terms of the topology of the CY, and experiments that measure the QCD axion mass thus constrain the topology. Our main result is summarized by Fig.~\ref{fig:h11predictions}: we present a distribution of the fraction of CYs in our ensemble that give rise to a detectable QCD axion in mass ranges covered by representative experiments. We now explain how this result was arrived at.

\textit{Pre-inflation relic abundance and fine tuning}---In this scenario, the QCD axion relic abundance is found by solving the homogeneous Klein-Gordon equation, which, at temperatures above the QCD crossover, is: $\ddot{\theta}+3H(T)\dot{\theta}+\frac{\chi(T)}{f_a^2}\sin\theta = 0$, where $H(T)$ is the Hubble rate given the standard model degrees of freedom~\cite{Saikawa:2018rcs}, and $\chi(T)$ is the QCD topological susceptibility~\cite{Borsanyi:2016ksw}, giving $m_a(T) = \sqrt{\chi(T)}/f_a$. For small $\theta_i$, the axion field begins to oscillate when $3H(T_{\rm osc})\approx m_a(T_{\rm osc})$, after which time the axion number density is conserved. For large $\theta_i$ the time of oscillations receives logarithmic corrections~\cite{Lyth:1991ub}, which can be accounted for by a correction factor $\mathcal{F}(\theta_i)$ to the relic density (which we compute following Refs.~\cite{Marsh:2015xka,Diez-Tejedor:2017ivd}):
\begin{equation}
    \rho_a \approx \frac{1}{2}m_a(T_0)m_a(T_{\rm osc}) f_a^2 \theta_i^2 \mathcal{F}(\theta_i) \frac{g_{S}(T_0)}{g_{S}(T_{\rm osc})}\left(\frac{T_0}{T_{\rm osc}}\right)^3\, ,
    \label{eqn:relic}
\end{equation}
with $g_{S}$ the number of relativistic degrees of freedom in the entropy, and $T_0$ the temperature of the Universe today. 

%%%%%%%%%%%%%%%
%%%%%%%%%%%%%

\begin{figure}
    \includegraphics[width=1.05\linewidth]{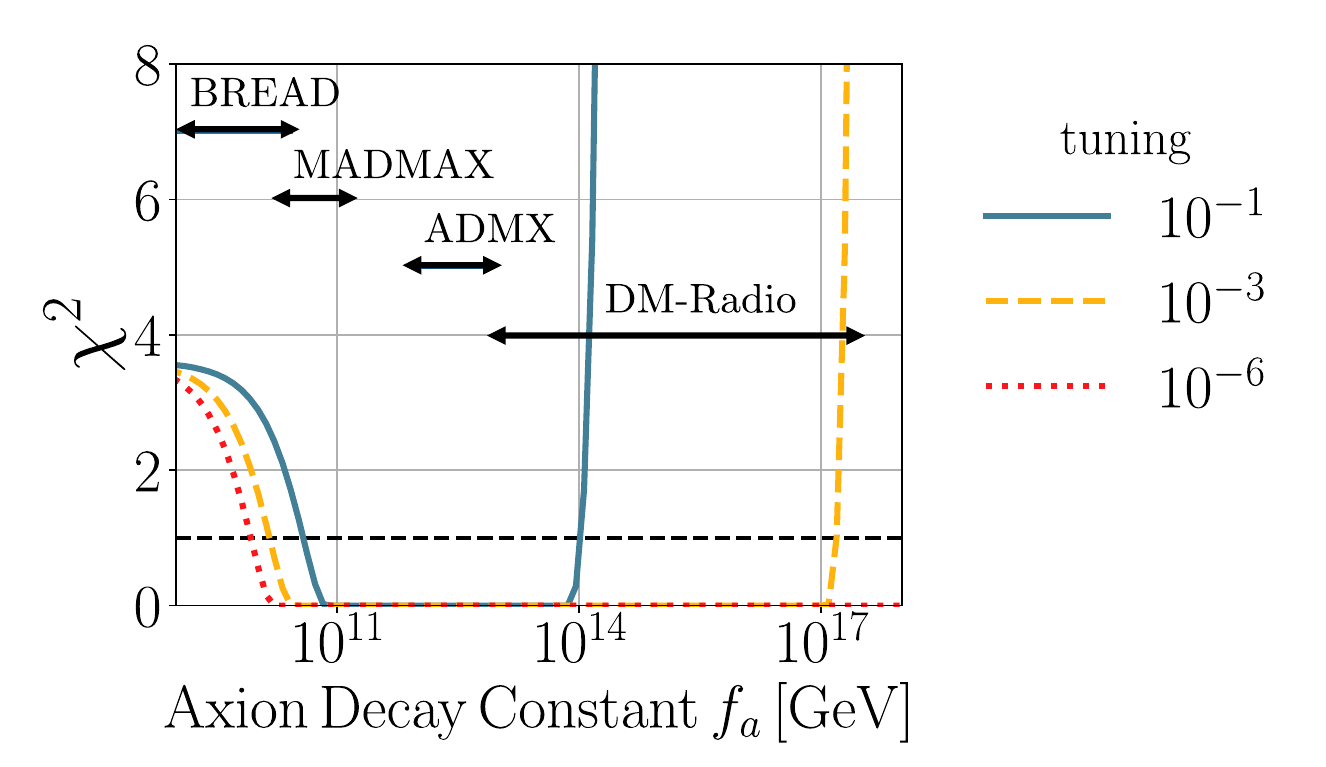}
    \caption{The minimum value of $\chi^2(f_a)$ from Eq.~\eqref{eqn:relic} and the \emph{Planck}~\cite{Planck:2018vyg} measurement approximated as a Gaussian likelihood. We allow fine tuning of the initial misalignment angle in the range $\theta_i\in [\epsilon,\pi-\epsilon]$. For moderate fine-tuning $\mathcal{O}(10^{-3})$ it is possible to accommodate large values of $f_a$, but values $f_a\lesssim 10^{10}\text{ GeV}$ are severely penalized even for very small $\epsilon$. We impose $\chi^2<1$ in our string theory ensemble to select models where the QCD axion can be all of the DM in the pre-inflation scenario with $\epsilon=10^{-2}$.}
    \label{fig:chisquared}
\end{figure}

In order for the QCD axion to explain the observed DM density, $\Omega_a h^2 =\rho_a/8.1\times 10^{-11}\text{ eV}^4\approx 0.12$, we impose a Gaussian likelihood for $\Omega_a h^2$ following Ref.~\cite{Planck:2018vyg}. Fixing $\theta_i\sim 1$, one obtains too large axion relic density at large $f_a$, and too small axion relic density at small $f_a$. We compute the range of allowed $f_a$ given a fine tuning measure on $\theta_i$. We do so by finding the range where $\chi_{\rm min}^2<1$ subject to the range $\theta_i\in [\epsilon,\pi-\epsilon]$ for given $\epsilon$. The result is shown in Fig.~\ref{fig:chisquared}. Allowing $\theta_i$ tuned small with $\epsilon\geq 10^{-3}$ allows $f_a\lesssim 10^{17}\text{ GeV}$. For small decay constants, however, even at this level of fine tuning the anharmonic corrections are only sufficient to allow the QCD axion to compose the entirety of the DM for $f_a\gtrsim 10^{10}\text{ GeV}$. As we will see, the fine tuning penalty in the pre-inflation scenario has a significant effect on conclusions related to the landscape.% (for an in depth statistical look at the pre-inflation scenario and resonator axion searches, see Ref.~\cite{Hoof:2018ieb}). \tkDM{Thinking about this, in a landscape context at 491, one could argue to allow much more fine tuning, since the staistics involve such large numbers of CY's?}
%%%%%%%%%%%%%
%%%%%%%%%%%%%%

 \textit{String theory setup}---Our basic setup is to compactify type IIB string theory on (an orientifold of) a CY threefold $X$, realized as a hypersurface in a toric variety (we will not construct explicit orientifolds, therefore tacitly assuming that one exists with $h^{1,1}_- = 0$). With the inclusion of fluxes in the internal manifold, the resulting four-dimensional scalar F-term potential can be characterized in terms of a K\"ahler potential $\mathcal{K}$ and superpotential $W$:
\begin{align}
    \mathcal{K} = -2\log \mathcal{V}, \ \ W = W_0 + \sum_\ell A_\ell e^{-2\pi q^i_\ell T_i}\, ,
\end{align}
where $\mathcal{V}=\mathrm{vol}(X)$, $W_0$ is the flux-induced Gukov-Vafa-Witten superpotential \cite{Gukov:1999ya}, the $A_\ell$ are the Pfaffian prefactors, which in general depend on the complex structure moduli, $q_a^i$ are instanton charges, and $i = 1, \ldots, h^{1,1}$. The fields appearing in the exponential are the complexified K\"ahler moduli, given by
\begin{align}
    T_j  &:= \tau_j + i \, \theta_j = \frac{1}{2} \int_{D_j} J \wedge J + i \int_{D_j} C_4, 
\end{align}
where $J$ is the K\"ahler form on $X$, $D_j$ is a four-cycle, and $C_4$ is the ten-dimensional Ramond-Ramond four-form potential. We will take $W_0 = A_\ell =1$ in this Letter and leave the effects of supersymmetry breaking and fermion zero-mode counting to future endeavors (the instanton actions are only logarithmically sensitive to $W_0$, see e.g. \cite{Sheridan:2024vtt}). Thus defined, the $\tau_i$ are the volumes of a basis of divisors (4-cycles), and $\theta_i$ are the axions in the same basis.

%The four-dimensional scalar potential is given by
%\begin{align}
%    V_F = e^K(K^{i\bar{j}} D_i W D_{\bar{j}} W - 3 |W|^2)
%    \label{eq:fterm}
%\end{align}
%where $K^{i\bar{j}}$ is the inverse K\"ahler metric determined by $\mathcal{K}$, and $D_i = \partial_i + \partial_i \mathcal{K}$.
We do not fashion explicit realizations of the standard model (SM) in this setup: rather, we consider toy models of the SM as stacks of D7-branes on various calibrated four-cycles in the CY. We do not engineer the chiral spectrum of the SM, but we impose on the geometry that the infrared gauge couplings of the SM are in the right range. Namely, the microscopic gauge coupling of an ${\rm SU}(N)$ gauge theory living on a stack of D7-branes wrapping a four-cycle $D$ is given by
\begin{align}
    \frac{1}{g_{\text{UV}}^2} \propto \mathrm{vol}(D).
\end{align}
Given a four-cycle volume $\mathrm{vol}(D)$, the IR gauge coupling depends on the $\beta$ function of the gauge theory. For QCD on a four-cycle $D_{\mathrm{QCD}}$, assuming no extra vectorlike pairs, low-scale supersymmetry implies $\mathrm{vol}(D_{\mathrm{QCD}}) \approx 25$, while high-scale (or no) supersymmetry implies $\mathrm{vol}(D_{\mathrm{QCD}}) \approx 40$. In this Letter, we impose $\mathrm{vol}(D_{\mathrm{QCD}}) =40$ as a benchmark value. 

The parameter space of 2- and 4-cycle sizes for a given CY $X$ is known as the K\"ahler cone, $K_X$. Without any sources of supersymmetry breaking, these parameters are massless moduli, which are experimentally excluded. Stabilizing these moduli is an important problem in string theory, and in this Letter we will make the assumption that the moduli can be stabilized by perturbative corrections to $\mathcal{K}$ at any point in $K_X$, or more specifically in some region where the gauge couplings take the correct values. The interplay of axion physics and more explicit moduli stabilization schemes has been studied in Refs.~\cite{Conlon:2006tq, Cicoli:2012sz, Cicoli:2013ana, Broeckel:2021dpz}.

We select points in $K_X$ via the following method: first, choose the point closest to the origin in $K_X$ such that that all 2-cycles have volume at least $1$ in string units. Then, choose a four-cycle $D_{\text{QCD}}$ on which to host a toy version of QCD. Perform a homogeneous rescaling of the K\"ahler parameters such that $\mathrm{vol}(D_{\mathrm{QCD}}) = 40$. It was observed in Ref.~\cite{Demirtas:2021gsq} that the resulting axion physics is not highly sensitive to angle within the K\"ahler cone, so we take the point chosen in this way as representative. 

Having chosen a point in moduli space, we compute the QCD axion decay constant. The axion potential is estimated by
\begin{align}
    V_{\mathrm{QCD}} &\approx \chi(T)U(\theta_{\rm QCD}) \nonumber \\+ &\sum_\ell 8\pi  \frac{\vec{q}_{\ell}\cdot \vec{\tau}}{\mathcal{V}^2} e^{-2\pi \vec{q}_\ell \cdot \vec{\tau}} [1-\cos(2 \pi q^i_\ell \theta_i)],
    \label{eq:axionpotential}
\end{align}
where $U(\theta)$ is the periodic potential generated by low-scale non-perturbative QCD effects, which is a cosine at high $T$~\cite{Gross:1980br,Wantz:2009it,Borsanyi:2016ksw,GrillidiCortona:2015jxo}. The kinetic term for the axions is given by
\begin{align}
    \mathcal{L}_{\mathrm{kin}} = -\frac{M_{\mathrm{pl}}}{2} K_{ij} \partial_\mu \theta^i \partial^\mu \theta^j\, ,
\end{align}
where $K_{ij} = 2 \, \partial_{T_i} \partial_{\bar{T}_{j}} \mathcal{K}$ and $M_{\mathrm{pl}}$ is the reduced Planck mass. We assume here that the relevant 4-cycles are not in highly-warped regions of the CY \cite{Martucci:2016pzt, Frey:2013bha, Hebecker:2018yxs}.
%We can now compute the decay constants of all the axions in this theory, following \cite{Glimmers}. The crux of the algorithm is to diagonalize the mass matrix in the approximation that $\Lambda_a^4 \gg \Lambda_{a+1}^4$ $\forall a$, using the Gram-Schmidt process.

The decay constant of the QCD axion, in the approximation that the instanton scales in \eqref{eq:axionpotential} are hierarchical, is~\cite{Gendler:2023kjt}:
\begin{align}
    f_{a} = \frac{M_{\mathrm{pl}}}{2\pi} \left[\sum_\ell q_{\ell n} (\mathbb{M}^{-1})^\ell_n \right]^{-1} \, ,
    \label{eq:qcddecay}
\end{align}
where the index $n$ denotes the cycle $D_{\mathrm{QCD}}$ and $\mathbb{M}$ is the change-of-basis matrix that takes the original charges in \eqref{eq:axionpotential} to the kinetic and approximate mass eigenbasis (see Ref.\cite{Gendler:2023kjt} for more detail). 

\textit{Axion experiments}---Proposed axion experiments will probe $10^{-11}\text{ eV}\lesssim m_a\lesssim 10^{-2} \text{ eV}$ for the QCD axion~\cite{Adams:2022pbo}, within the limits imposed by astrophysics~\cite{Arvanitaki:2014wva,Stott:2018opm,Hoof:2024quk,Burrows:1988ah,Raffelt:2006cw,Lella:2023bfb,Caputo:2024oqc} and $f_a<M_{\mathrm{pl}}$. %from the lower mass limit imposed by $f_a<M_{pl}$ \nvg{what experiments actually get down to this lower limit?} \tkDM{CASPEr supposedly} and black hole superradiance~\cite{Arvanitaki:2014wva,Stott:2018opm,Hoof:2024quk}, to the upper limit imposed by the duration of the SN1987A neutrino burst~\cite{Burrows:1988ah,Raffelt:2006cw,Lella:2023bfb}. 
In this Letter, we consider four representative `haloscope' axion DM experiments and their target mass ranges: DMRadio~\cite{DMRadio:2022pkf}, ADMX~\cite{Stern:2016bbw}; MADMAX~\cite{Beurthey:2020yuq} and BREAD~\cite{BREAD:2021tpx} (the implications for other experiments and proposals, e.g. Refs.~\cite{Horns:2012jf,Budker:2013hfa,Marsh:2018dlj,Lawson:2019brd,Alesini:2023qed} can be inferred from the mass range). Importantly, all of these projections assume the QCD axion to be all of the local DM density. We also consider the ``helioscope" IAXO, which can detect and measure the mass of QCD axion models with large $C$ in the range $10^{-3}\text{ eV}\lesssim m_a\lesssim 10^{-1} \text{ eV}$~\cite{Dafni:2018tvj,Hoof:2021mld}. IAXO detects axions produced in the sun, and thus does not depend on the relic abundance, but for the QCD axion to be detectable requires an increased $\mathcal{C}$~\cite{Dafni:2018tvj,Hoof:2021mld}, which can be achieved via anomalies~\cite{DiLuzio:2016sbl,DiLuzio:2020wdo,Plakkot:2021xyx} (with the QCD axion interacting only via the D-brane Chern-Simons action, plus IR mixing with the pion, we have $C\approx\pm 1-0.87$~\cite{Gendler:2023kjt}).

% DM-radio: $2\times 10^{-11}\text{ eV}\leq m_a\leq 8\times 10^{-7} \text{ eV}$
% ADMX: $7\times 10^{-7}\text{ eV}\leq m_a\leq 5\times 10^{-6} \text{ eV}$
% MADMAX: $5\times 10^{-5}\text{ eV}\leq m_a\leq 2\times 10^{-4} \text{ eV}$
% BREAD: $2\times 10^{-4}\text{ eV}\leq m_a \leq 1 \text{ eV}$

% , which in the pre-inflation scenario is $\rho_{\rm DM}\approx 0.4\text{ GeV cm}^{-3}$~\cite{deSalas:2020hbh} (c.f. post-inflation~\cite{Eggemeier:2022hqa})

%\nvg{review of resonator experiments, plots}
%\begin{itemize}
%    \item chi squared vs f (put in regions of f for each experiment (stacked arrows?))
%    \item histograms of fractions with h11 on x axis for each experiment
%    \item mass histogram KDEs for each h11 for QCD
%    \item helioscopes plot (like second plot but just for iaxo bin) (put this in main plot, just as a dotted curve)
%\end{itemize}

%The main goal of this work is to provide quantitative expectations for $h^{1,1}$ of Calabi-Yau compactifications of type IIB string theory in the event of a detection of the QCD axion as dark matter. Specifically, we focus on resonator-based experiments, which rely on a background of fixed-mass axion dark matter in order to see a signal. The resonator experiments we analyze here are BREAD \cite{}, MADMAX \cite{}, ADMX \cite{}, and DM-Radio \cite{}. \nvg{do we need to say more about how these experiments actually work?}

%%%%%%%%%%%%%
\begin{figure}
    \includegraphics[width=1\linewidth]{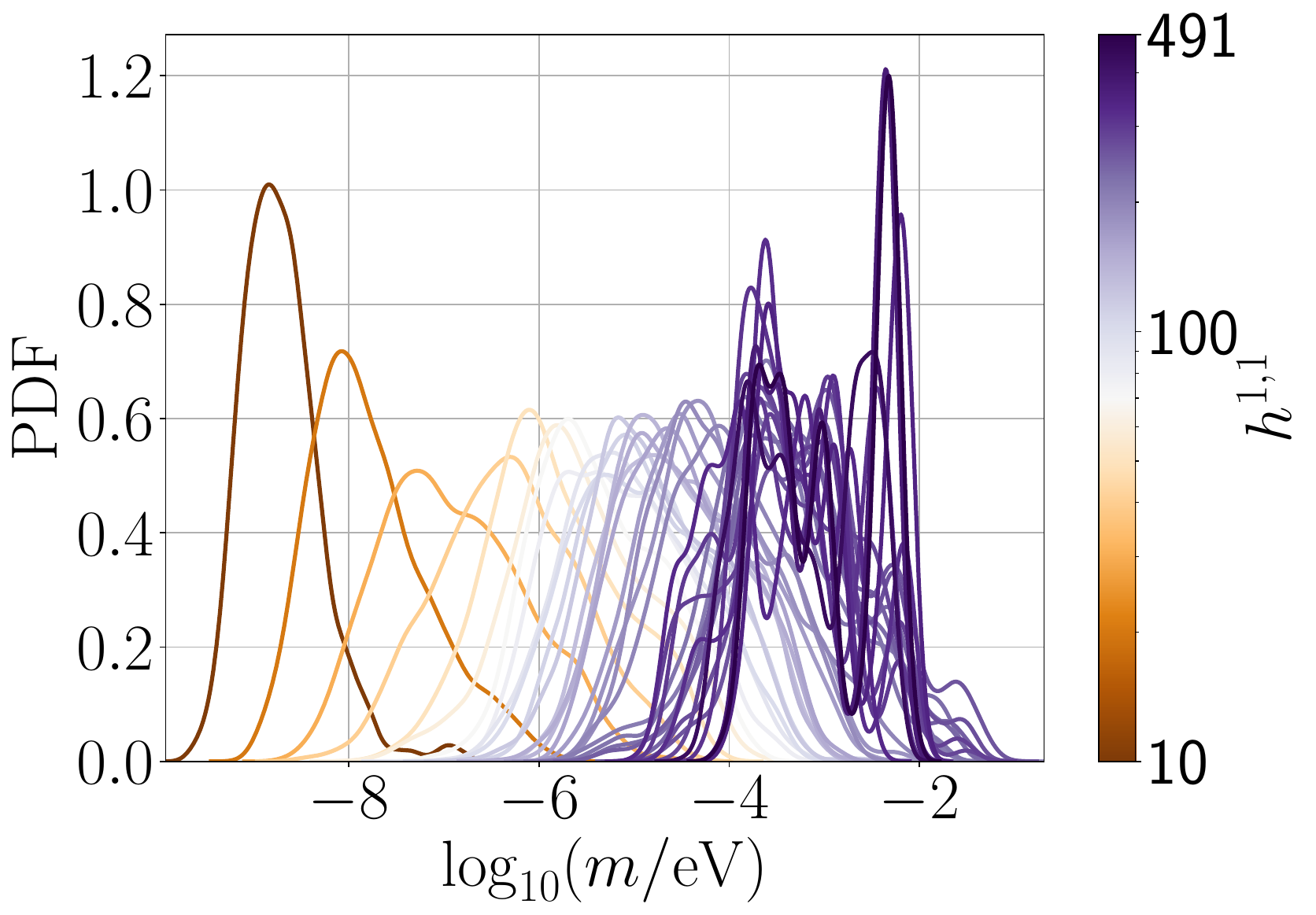}
    \caption{Kernel density estimates of the QCD axion mass for each value of $h^{1,1}$ in the dataset.}
    \label{fig:masses}
\end{figure}
%%%%%%%%%%%%%%

 \textit{Results}---An important lesson that can be drawn from numerous studies of string axions~\cite{Demirtas:2021gsq, Mehta:2020kwu,Mehta:2021pwf, Gendler:2023kjt, Demirtas:2018akl, Halverson:2019cmy} is that axion decay constants (and thus QCD axion mass) are highly correlated with $h^{1,1}$. The axion experiments given above are sensitive to the QCD axion mass in different ranges. Therefore, a detection by any one of these experiments allows us to make an inference on $h^{1,1}$ in the landscape of string theory compactifications. We now make this correlation between $h^{1,1}$ and experiments concrete. 

To this end, we analyzed an ensemble of $185{,}965$ CY threefolds obtained as hypersurfaces in toric varieties. These manifolds are explicitly made via Batyrev's construction~\cite{batyrev1993dual} by triangulating polytopes in the Kreuzer-Skarke database~\cite{Kreuzer:2000xy} using \textsc{CYTools}~\cite{CYtools}. Our scan consisted of at most $100$ randomly selected polytopes per $h^{1,1}$, sampled in steps of $10$ starting at $h^{1,1}=10$, with the addition of the single polytope with $h^{1,1}=491$ (the largest in the set). For each polytope, a set of  triangulations was obtained using the \texttt{random\_triangulations\_fast} method in \textsc{CYTools}. The number of triangulations sampled per polytope was chosen such that the total number of triangulations per $h^{1,1}$ was $1000$.  As pointed out in Ref.~\cite{Demirtas:2020dbm}, this sampling method does not necessarily produce a fair sample on the space of triangulations. Sampling triangulations and inequivalent CYs are difficult problems that we are currently investigating, but we do not expect them to qualitatively change the picture presented here \cite{MacFadden:2024him}. To mitigate these effects, we do sample fairly over the set of polytopes, which is possible because this set is finite. %Differences in sampling methods is being investigated in \cite{pipeline}.

For each Calabi-Yau threefold in our ensemble, we randomly choose at most five four-cycles to host QCD, as well as a point in moduli space, as described in the previous section. Once these data are specified, we read off the QCD axion decay constant using \eqref{eq:qcddecay}, and bin by experiment. %We then compare these results with the ranges of sensitivity for the various resonator experiments described above.

Our results are shown in Fig~\ref{fig:h11predictions}, giving the fraction of CYs in the ensemble at each $h^{1,1}$ where the QCD axion falls in the range of each experiment. Because of the correlation of $f_{a}$ with $h^{1,1}$, detection of the QCD axion as DM by any experiment would favour particular values of $h^{1,1}$ in the ensemble. %The main point is that the fraction of Calabi-Yau threefolds to which a given resonator experiment is sensitive to depends on the experiment. Furthermore, a detection by a given experiment of QCD axion as all of the dark matter would indicate a preference for a different range of $h^{1,1}$ of the underlying Calabi-Yau manifold. In particular, the distributions for DM Radio, ADMX, MADMAX, and BREAD peak at $h^{1,1}=26.4, \ 72.3, \ 208.7,$ and $283.6,$ respectively.

The most intuitive way to understand these results is to note the correlation of the QCD axion mass with $h^{1,1}$ in this landscape. To this end, in Fig.~\ref{fig:masses}, we show the distribution of masses (using kernel density estimation) for each value of $h^{1,1}$ in our dataset. Figure ~\ref{fig:masses} includes every instance of the QCD axion, rather than restricting to instances in which the QCD axion can compose all of the dark matter.

To investigate the effect on $f_a$ of taking the QCD divisor with volume 40 found as specified above, we used Markov chain Monte Carlo sampling~\cite{2013PASP..125..306F} of the Weil-Petersson prior within the stretched K\"{a}hler cone with a Gaussian likelihood for $\mathrm{vol}(D_{\mathrm{QCD}}) = 40 \pm 1$ on a single CY with $h^{1,1}=10$ (a more complete study of the landscape using this method is beyond the scope of the present Letter, but is under investigation). We found that the resulting spread in $f_a$ is $\mathcal{O}(1)$. We also ran our analysis imposing $\mathrm{vol}(D_{\mathrm{QCD}}) =25$ and found that our results do not change significantly.

 \textit{Discussion}---We have demonstrated the immense power of experiments to cut down the string landscape by measuring the QCD axion mass and thus making an inference on $h^{1,1}$. Though the counting measure on the set of toric hypersurface CYs is not known, the polytope with the overwhelming majority of (possibly equivalent) triangulations has $h^{1,1}=491$~\cite{Demirtas:2020dbm}. In our scans and previous work~\cite{Gendler:2023kjt} we found that this case has mean decay constant $2.6\times 10^9\text{ GeV}$, which cannot give all of the DM without severe fine tuning $\epsilon\ll 10^{-6}$. Nonetheless, around 20\% of models with $h^{1,1}=491$ lie in the untuned BREAD and MADMAX sensitivity regions. On the other hand, if $\mathcal{C}$ is large, IAXO offers sensitivity to around 50\% of the $h^{1,1}=491$ samples (from a string theory perspective, this mass range was proposed in \cite{Gendler:2024gdo}).

Though this Letter was carried out in the context of type IIB string theory compactified on Calabi-Yau threefold orientifolds, we expect that  qualitatively, the results here extend to other corners of the string landscape. The structures that drive correlations between the QCD axion mass and the hodge number $h^{1,1}$ are geometric, and are therefore likely to induce similar behavior in compactifications of other string theories. Other classes of compactification manifolds also exhibit similar features (see Ref. \cite{Halverson:2019cmy} for work involving F-theory on Calabi-Yau fourfolds).

We have considered only closed string axions, which are necessarily in the pre-inflation regime~\cite{Reece:2024wrn}, with the inflationary scale below the compactification scale (if this is not the case, the early Universe would be highly exotic and almost impossible to study analytically, although see Refs.~\cite{Greene:1989ya,March-Russell:2021zfq,Benabou:2023npn}). In string theory, the alternative post-inflation scenario only seems possible if the QCD axion is from the \emph{open string} sector, which can occur when the Standard Model is realised by D3 branes at singularities (e.g. Refs.~\cite{Cicoli:2012vw,Cicoli:2013mpa,Cicoli:2021dhg}), which we have not considered in this Letter. It is possible, however, that experiments and astrophysics could determine the axion to be in the post-inflation scenario by detecting the ``miniclusters" predicted by cosmic string decay~\cite{Beurthey:2020yuq,BREAD:2021tpx,Schutte-Engel:2021bqm,Fairbairn:2017dmf,OHare:2017yze,Edwards:2020afl,Ellis:2022grh,Witte:2022cjj,OHare:2023rtm,Gorghetto:2024vnp}. Such a detection would have very profound implications for string theory.  Further study of post-inflation and open string QCD axions in string theory is warranted, in particular in the KS database at large $h^{1,1}$. Detection of the QCD axion, either as dark matter or through a helioscope, may provide an indication of our Universe's location in the string theory landscape.

%With arbitrary fine tuning, 491 gives you something in BREAD. 491 could be detected by IAXO if $C$ is large. In this set up, the QCD axion is a \emph{closed string axion}, i.e. it is part of a complexified K\"{a}hler modulus $T=\tau+i\theta$ and is \emph{not} the phase of a matter field, $\Phi$. This is in contrast to field theory models for the axion, and in particular precludes the `post-inflation' scenario. Closed string axions are necessarily in the pre-inflation scenario since the UV scale is the compactification scale, and if this scale is below the inflationary scale then the four dimensional effective theory is not valid, making predictive physics difficult (although see Refs.~\cite{Greene:1989ya,March-Russell:2021zfq,Benabou:2023npn}).

\textit{Acknowledgments}---We thank Liam McAllister and Jakob Moritz for discussions and helpful comments on a draft. We thank Béla Majorovits for discussions that inspired this work. We thank Edward Witten for some reassuring correspondence. DJEM is supported by an Ernest Rutherford Fellowship from the STFC, Grant No.~ST/T004037/1 and by a Leverhulme Trust Research Project (RPG-2022-145). The work of NG was supported in part by a grant from the Simons Foundation (602883,CV), the DellaPietra Foundation, and by the NSF grant PHY-2013858. We would like to thank the Harvard Swampland Initiative program, where this work was initiated.

\bibliography{refs}

\end{document}